\documentclass[]{ceurart}

\newcolumntype{H}{>{\setbox0=\hbox\bgroup}c<{\egroup}@{}}

\begin{document}



\title{NeuralSearchX: Serving a Multi-billion-parameter Reranker for Multilingual Metasearch at a Low Cost
}
\author[]{Thales Sales Almeida*}[%
email=thalesrogerio@gmail.com,
]
\author[]{Thiago Laitz*}[]
\author[]{João Seródio*}[]
\author[]{Luiz Henrique Bonifacio}[]
\author[]{Roberto Lotufo}[]
\author[]{Rodrigo Nogueira}[]

\address[]{NeuralMind, Brazil}

\copyrightyear{2022}
\copyrightclause{Copyright for this paper by its authors. Use permitted under Creative Commons License Attribution 4.0 International (CC BY 4.0).}

\conference{DESIRES 2022 -- 3rd International Conference on Design of Experimental Search \& Information REtrieval Systems, 30-31 August 2022, San Jose, CA, USA}

\begin{abstract}
  The widespread availability of search API's (both free and commercial) brings the promise of increased coverage and quality of search results for metasearch engines, while decreasing the maintenance costs of the crawling and indexing infrastructures. However, merging strategies frequently comprise complex pipelines that require careful tuning, which is often overlooked in the literature. In this work, we describe NeuralSearchX, a metasearch engine based on a multi-purpose large reranking model to merge results and highlight sentences. Due to the homogeneity of our architecture, we could focus our optimization efforts on a single component.
  We compare our system with Microsoft's Biomedical Search and show that our design choices led to a much cost-effective system with competitive QPS while having close to state-of-the-art results on a wide range of public benchmarks. Human evaluation on two domain-specific tasks shows that our retrieval system outperformed Google API by a large margin in terms of nDCG@10 scores. By describing our architecture and implementation in detail, we hope that the community will build on our design choices. The system is available at \url{https://neuralsearchx.neuralmind.ai}.\nonumnote{*Equal contribution.}
  
\end{abstract}


\maketitle

\section{Introduction}

Metasearch engines provide a unified interface for searching and aggregating results from different sources. They take advantage of existing search engines to increase the diversity of results while decreasing maintenance costs for crawling and indexing infrastructures.

This approach, however, comes with a unique set of challenges: many times metasearch engines choose to consult only sources that might be relevant to a given query, since searching all available collections may be unfeasible due to resource constraints, such as limited bandwidth. For example, given a medical-related query, a metasearch engine could choose to consult only search engines that are capable of returning medical-related content. Therefore, a metasearch engine must represent the capabilities and type of data that each source search engine provides, i.e., the representation problem. When receiving a query, the metasearch engine must be able to select the appropriate sources, i.e., the selection problem. Furthermore, a metasearch engine is also responsible for generating a unified list of results from the multiples lists that it retrieves from, i.e., the merging problem.

In this work, we introduce NeuralSearchX, a multi-stage metasearch engine that takes a different approach to metasearch: to improve the quality of results we retrieve mainly from well-established general-purpose engines, such as Google and Bing, as well as curated sparse and dense indexes. Therefore, we avoid the representation and selection problem as we can simply search all our bases for any given query. To address the merging problem, NeuralSearchX leverages state-of-the-art Transformer~\cite{transformers} models as a merging strategy. Such models recently showed promising zero-shot capabilities~\cite{rosa2022scaling}, i.e., they can perform well in unseen domains and therefore are suitable for general web search.

Our approach does not require any specific knowledge from the source collections since it ranks documents purely by content. We show in the experimental section that our system outperforms well-established search engines even without extra information from the document or its source.

NeuralSearchX also takes advantage of low-reliability cloud infrastructure in its deployment in order to use high computing power while keeping a low overall cost.

Our contributions are the following:
\begin{itemize}
    \item We propose zero-shot Transformer rerankers as an effective merging strategy for metasearch engines. We demonstrate the effectiveness of our reranking pipeline on various public datasets as well as human evaluations;
    \item By using low-reliability infrastructure, we show that it is possible to use a model with billions of parameters in a production environment under a reasonable budget, thus allowing us to deploy a highly effective search engine; 
\end{itemize}

\section{Related Work}

In this section, we first discuss metasearch engines and their main challenges. Then we provide an overview of multistage retrieval pipelines.

\subsection{Metasearch Engines}

Metasearch engines typically support hundreds or even thousands of search engines and therefore demand highly elaborated solutions for the representation, selection and merging problems~\cite{meta1liu2007allinonenews,meta2gauch1996information,meta3gulli2005building}.

Metasearch engines traditionally handle the collection representation problem by maintaining representational sets that contain pertinent information for each searchable collection. Such sets can be created manually~\cite{manber1997search} but are usually generated automatically due to scaling reasons. Various methods were proposed to generate such sets~\cite{coop1d2004collection,coop2xu1998effective,coop3gravano1997starts,noncoop1nottelmann2003evaluating,noncoop2si2004unified}. For example, Query-Based Sampling (QBS) \cite{callan2001query} generates the representation set by sampling documents from the collection using short queries to infer the collection content.

For the selection problem, existing solutions often calculate similarity metrics between the query and the representational sets of each collection~\cite{lexicon1gravano1997querying,lexicon2baumgarten1999probabilistic}, and then select the collections with the best scores. Other approaches include ranking collections based on the contents of the top returned documents~\cite{tech_exp1rasolofo2001approaches}, or estimating the probability that a given collection has at least one relevant document~\cite{tech_exp2larson2002logistic}.

Traditional merging strategies are based on scoring functions that take into account textual content and metadata retrieved from the source search engines. For example, CORI~\cite{cori1callan2002distributed,cori2callan1995searching} assigns a belief score for each collection based on a Bayesian inference network. The belief score is then used in the scoring function to determine the score of each retrieved document in that collection. SAFE~\cite{Safe1shokouhi2009robust} aggregates all retrieved documents as well as samples available from the source collections, and fits a statistical model to predict the scores. More recently, \citet{merg1vijaya2016artificial} proposed the use of a neural network as a merging strategy by providing the model with numerous corpus and document-level statistics as input in order to compute the relevancy of the document. Note that this approach is different from the one implemented by NeuralSearchX since our models score relevance based on the document content and do not require any extra information. Finally, \citet{metricsMetalu2005evaluation} make a comparative analysis of a wide variety of merging strategies and demonstrates that a well-performing merging strategy is fundamental to achieve adequate results.

\subsection{Multistage Ranking}

A multistage ranking retrieval pipeline is a common retrieval pipeline that uses multiple consecutive refinement steps to return the final results. This method was first referenced back in 2006 at \citet{matveeva2006high}. In 2010 it was revealed that the Bing search engine used multistage ranking to deliver its results, in the next few years, a handful of notorious search systems also related to multistage ranking \cite{cambazoglu2010early,huang2020embedding} but with different numbers and types of stages.

In the last decade, we have seen an increasing dedication to improving the quality of the results by proposing innovations on different stages of the retrieval pipeline~\cite{liu2017cascade,liu2021pre,zhang2020machine}. We also saw a special focus on optimizing search systems for more specific and challenging domains. In this context, Covidex~\cite{zhang2020covidex} and Biomedsearch~\cite{Wang_2021} have recently been proposed search engines that focus on providing scientific information in the medical domain. Both use a multistage ranking similar to that of NeuralSearchX.

\section{Our Solution}

\begin{figure*}
    \centering
    \includegraphics[width=1\linewidth]{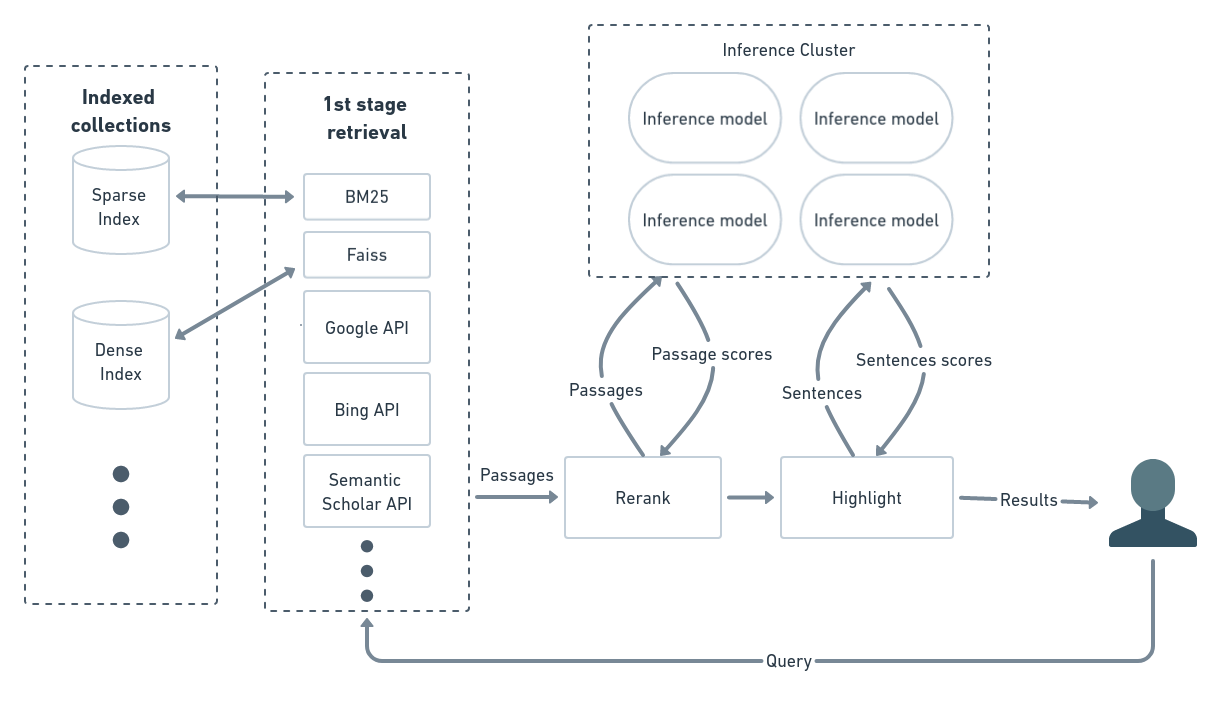}
    \caption{Retrieval architecture of NeuralSearchX. Note that reranking and highlighting are performed by the same neural model, which decreases hardware idle time.}
    \label{fig:gen_pipe}
\end{figure*}

NeuralSearchX is a multi-stage metasearch engine: the first stage consists of a candidate document retrieval step in which documents are retrieved using federated search over web search engines along with sparse and dense retrieval from private collections. The second stage consists of a reranking step of the previously selected documents with a neural model. The model scores each document by how relevant it is for a given query. The best scored documents are then sent to a highlighter model that estimates and selects the most relevant sentences from each document. Finally the documents are formatted and returned to the end user. Figure~\ref{fig:gen_pipe} illustrates the described pipeline.

\subsection{1st stage: Candidate Document Retrieval}

For the first stage, we use both sparse and dense retrieval methods provided by the Pyserini's~\cite{lin2021pyserini} library. We choose to use Pyserini to take advantage of its support for evaluation on well-established IR benchmarks.

Before indexing, we split the contents of long documents over multiple windows. We do this by splitting the original content in words and grouping them according to a certain window size and stride. Typically we use windows of 150 words with a stride of 75 words. We use 150 words because it translates to about 250 tokens, which is close to the max length of 256 tokens used to train our models. This process is done to improve the efficiency of the Transformer model used in the next stages, since it has a quadratic cost over the length of the input sequence. It also improves its effectiveness since this window length is close to the average length of the texts that the model was finetuned on, i.e., the passages from the MS MARCO dataset.

Furthermore, NeuralSearchX also leverages third-party APIs, such as Bing, Google and Semantic Scholar, to retrieve documents from the web. We assume that the snippets returned by such APIs are representative enough of the original page; therefore we use the provided snippets as the document content. A single search can use multiple of such sources to create a diverse set of documents. Since all the retrievals are independent, we can run them in parallel, therefore not creating a significant overhead in the process and maintaining an adequate latency.

\subsection{2nd stage: Reranking}

After the candidate document retrieval, the next step is to merge all the candidates in a single list, and then rank such documents so that the most relevant ones are on the top of the results list. To do that, NeuralSearchX uses one of the following rerankers:
\begin{itemize}
    \item A mMiniLM reranker introduced by \citet{translate_DBLP:journals/corr/abs-2108-13897}. It has 107 million parameters and was trained on the English and Portuguese subsets of the mMARCO~\cite{translate_DBLP:journals/corr/abs-2108-13897} dataset, a machine translated version of the MS MARCO dataset~\cite{bajaj2016ms}. This distilled model has been shown to perform better than models with an order of magnitude more parameters \cite{rosa2022scaling}.
    \item A mT5 model with 3.7 billion parameters. The model is based on the T5 model and was proposed by \citet{nogueira2020document}. It was recently shown that this model yields SOTA results in zero-shot scenarios~\cite{rosa2022scaling}. Here we use its multilingual version that was trained on all languages of mMARCO.
\end{itemize}

The reranker computes the relevancy of each document for a given query. After all documents are scored, the list of results is reordered according to these scores. 

\subsection{3rd stage: Highlighting}

With the results ordered by their relevance, NeuralSearchX takes the first ten ranked documents and performs a highlighting step, whose goal is to take long documents and select the most important sentences to show the user, therefore reducing the amount of text that the user needs to read. Note, however, that this step is not performed when the document text is already a snippet (such as those from Google and Bing), because the text in this case is already of an appropriate length. We begin by splitting the document into sentences and compute the relevance of each sentence to the query. Since the task is still the same as in the previous stage (only the text is shorter), we can leverage the same model and hardware used in the ranking stage. For each document, the two sentences with the highest scores are selected and then shown in the user interface.

\subsection{Tools and Infrastructure}

Like most modern search engines, NeuralSearchX was built in a cloud environment to take advantage of its distributed computing capabilities. We made sure however that our system was cloud agnostic, meaning it is easily adaptable to any major cloud provider. At the time of writing this paper, we had already deployed the system on Google Cloud and Microsoft Azure.

NeuralSearchX uses a frontend in a Single-page Application (SPA) model, and is built with React, a well-established javascript library for the creation of web pages with dynamic content. The webpage is served through an Nginx web server that runs on a dedicated machine. The same Nginx web server also acts as a load balancer for the requests to the API.

Our API is built using FastAPI, a Python framework for the construction of APIs based on the Starlette toolkit. FastAPI is capable of creating very well-performing and low-latency APIs. The API is responsible for handling document retrieval using both Pyserini's retrieval methods and through http requests to third-party APIs. It also handles basic parsing and formatting of documents.

\subsection{Low-reliability Infrastructure}

In recent years, we have seen increasing improvements delivered by neural models for information retrieval, by far surpassing traditional search methods with respect to the quality of results~\cite{lin2021pretrained}. However, such models come with a major drawback: they are much more computationally expensive than the traditional non-neural search. 

Due to this high computational cost, such models either have a high latency on modest hardware or demand costly hardware (mainly GPUs) to perform computations in reasonable time, making such methods less attractive for commercial search engines.

In the last couple of years, we also saw a common trend among the major cloud providers in the form of spot resources. Such an infrastructure is cheap but can be taken down by the provider with little to no warning. The use of low-reliability infrastructure was explored for training huge models~\cite{bottou2018optimization,dean2012large,yang2021scheduling} but little work has been published on its use in production systems, especially those that demand high computing power, such as systems powered by deep learning models. NeuralSearchX takes advantage of spot infrastructure to make the deployment of state-of-the-art models at a feasible price, allowing for both low query latency and a competitive low-cost infrastructure. Our work shows a practical use case in which spot instances can be used with success in a deployment scenario. 

Furthermore, in our experience, the spot resources that we use (mainly machines with T4 gpus) are not evicted often: in a period of three months, our available hardware was below 80\% of our total capacity only once. However, the time to start serving a model in a new machine can be up to 20 minutes, accounting for both the time to create a new container and loading the necessary models onto the GPU.

\subsection{Service orchestration}

The major drawback of using spot hardware is the lack of reliability, since resources can be deallocated at any given time. To mitigate this problem, we deploy our spot resources under a Kubernetes cluster that handles the allocation of spot machines. If a machine is taken down, Kubernetes promptly starts the process to allocate a new one with the equivalent services.

Being prepared to recover from resource eviction is essential when using spot machines. Depending on the scale and characteristics of the system in question, additional mechanisms may be necessary. For example, a hybrid scheme using both spot and non-spot resources would provide a reliable minimum computation while still achieving an adequate average capacity.

The most expensive part of our pipeline is the reranking and highlighting stages as they use modern GPUs to do hundreds of inference passes on a large neural network in under a few seconds. Such resources are deployed on a low-reliability infrastructure in a Kubernetes cluster. Currently we use a managed cluster in Azure Kubernetes Service (AKS), but this could be replaced with any other managed Kubernetes service from the major cloud providers. All neural models are served using TorchServe, a performant and reliant solution for the deployment of Pytorch~\cite{paszke2019pytorch} models. Each model has its own TorchServe image, which is deployed in pods that run on nodes of the Kubernetes cluster. This method of provisioning the reranker service allows us to modify and adapt the system in a seamless way. For example, we were able to deploy a model with 3.7 billion parameters by creating a torchserve image with the respective model and then deploying it under 10 machines with T4 GPUs. The API would simply separate the passages to be scored in 10 requests, which will then be distributed between the 10 nodes running the model service and return the results of the inference.

\section{Experiments and Results}

Two fundamental factors that determine the usefulness of a retrieval system are query latency and quality of results. In this section we analyze both of these aspects in NeuralSearchX, as well as its serving cost.

\subsection{Efficiency}

To evaluate the efficiency of our system we measure the queries per second (QPS) that our system can handle and compare it with its monthly cost. We also compare our results with those presented by \citet{Wang_2021}, who proposes a search engine with a similar pipeline.

We use Locust,\footnote{\url{https://locust.io/}} a Python framework for load testing, to measure the system QPS. We evaluate the efficiency of our system in the MS MARCO dataset with a standard BM25 for document retrieval. We also evaluate our search in the general web using third-party APIs. The simulated users from Locust were configured with the following characteristics:

\begin{enumerate}
    \item Each simulated user sends a random query from the query set.
    \item After each request, the simulated user waits 1 to 15 seconds to issue another query.
\end{enumerate}

Before the load test, we make sure that all relevant components are deployed under the same private network and were already deployed under the Kubernetes cluster, so we evaluate the system in its peak efficiency.

In the MS MARCO load test, we used the MS MARCO passage dev topics as our query set. The system was tested for 15 minutes. For the Web load test, we used a custom query set containing English and Portuguese queries. For this test the system was tested for 3 minutes. The results of the load tests for both models are shown in Table~\ref{tab:latency_results}. The column ``90\%' refers to the query latency of the 90th percentile.

\begin{table}[]
\caption{Overall QPS and infrastructure cost (in USD). All costs were based on Azure's pricing.}
\label{tab:QPS_and_Cost}
\begin{tabular}{lr|rrr}
\toprule
 &  & \multicolumn{3}{c}{Costs (USD)} \\
Engine & QPS & Initial Retrieval & Rerank+Highlight & Total \\ \midrule
NSX (mMiniLM) & 5.2 & 200 & 1.8k (2 V100's) & 2k \\
NSX (mT5) & 3.3 & 200 & 2.2k (10 T4's) & 2.4k \\ \midrule
MS Biomedsearch~\cite{Wang_2021} & 4 & 5k & 24k & 29k \\ \bottomrule
\end{tabular}
\end{table}

\begin{table}[]
\caption{Results of the efficiency test (QPS and seconds/query). The system that uses the mMiniLM model uses 2 V100 GPUs to perform inference, while the mT5 system uses 10 T4 GPUs.}
\label{tab:latency_results}
\begin{tabular}{l|rrrrr|rrrrr}
\toprule
 & \multicolumn{5}{c|}{MS Marco} & \multicolumn{5}{c}{Web Search} \\
Engine & QPS & 90\% & mean & min & max & QPS & 90\% & mean & min & max \\ \midrule
NSX (mMiniLM) & 5.2 & 1.904 & 1.385 & 0.502 & 9.430 & 4.8 & 0.910 & 0.670 & 0.379 & 1.408 \\
NSX (mT5) & 3.3 & 13.000 & 6.039 & 1.232 & 30.873 & 4.2 & 2.200 & 1.581 & 0.875 & 6.902 \\ \bottomrule
\end{tabular}
\end{table}

The projected monthly cost of our system is shown in Table \ref{tab:QPS_and_Cost}. In comparison to MS Biomedical Search Engine, we were able to achieve a similar QPS but with a much lower overall cost. The main reason for this is the ability to use spot instances for GPU-intensive services, as discussed previously.

\subsection{Evaluation on Public Datasets}

To evaluate search systems, numerous document collections were created and made publicly available. To measure the effectiveness of NeuralSearchX, we indexed two of such document collections and their respective automatic translations to Portuguese:

\begin{enumerate}
    \item MS MARCO~\cite{bajaj2016ms}: A large scale dataset introduced in 2016 comprising more than 8 million passages and 500 thousand queries issued to the Bing search engine and their respective relevant passages selected by humans. Our models were trained in query and document pairs from this dataset.
    \item Robust04~\cite{voorhees2003overview}: A benchmark that contains half million documents from the news domain and a large amount of annotations for 250 queries.
\end{enumerate}

\begin{table*}[]
\caption{Results on public benchmarks (MARCO and Robust04) and manual evaluations in different domains (Legal and Business). Results marked with $*$ use only 100 retrieved documents.}
\label{tab:all_effectiveness_results}
\begin{tabular}{HlccHcHccHHcc}
\toprule
 &  & \multicolumn{2}{c}{\textbf{MARCO}} & & \multicolumn{3}{c}{\textbf{Robust04}} & \multicolumn{1}{c}{\textbf{Legal}} & & & \multicolumn{2}{c}{\textbf{Business}} \\
 & & \multicolumn{1}{c}{\textbf{en}} & \multicolumn{1}{c}{\textbf{pt}} & & \multicolumn{1}{c}{\textbf{en}} & & \multicolumn{1}{c}{\textbf{pt}} & \multicolumn{1}{c}{\textbf{pt}} & & & \multicolumn{2}{c}{\textbf{pt}} \\
 &  & \scriptsize\textbf{MRR@10} & \scriptsize\textbf{MRR@10} & \textbf{MAP} & \scriptsize\textbf{nDCG@20} & \textbf{MAP} & \scriptsize\textbf{nDCG@20} & \scriptsize\textbf{nDCG@10} & \textbf{P@10} & \textbf{R@10} & \scriptsize\textbf{MRR@10} & \scriptsize\textbf{nDCG@3} \\
 \midrule
(1) & Google API & - & - & - & - & - & - & 0.6219 & 0.4500 & 0.4801 & - & 0.7805 \\
(2) & BM25 & 0.1840 & 0.1520 & 0.2531 & 0.4240 & 0.2307 & 0.3893 & - & - & - & 0.1211 & 0.5678 \\
(3) & NSX (mMiniLM) & \textbf{0.3749} & \textbf{0.3184} & 0.2419 & 0.4570 & 0.2085 & 0.3966 & \textbf{0.7063} & \textbf{0.5360} & \textbf{0.5273} & 0.1776 & - \\
(4) & NSX (mT5) & 0.3553$^*$ & 0.3036$^*$ & \textbf{0.2631} & \textbf{0.4964} & \textbf{0.2327} & \textbf{0.4464} & - & - & - & \textbf{0.2070} & \textbf{0.8774} \\
\bottomrule
\end{tabular}
\end{table*}

As mentioned previously, NeuralSearchX is capable of answering queries in languages other than English, since it uses multilingual models to rank the retrieved documents. To evaluate our system in other languages, we used the translated versions of MS MARCO and Robust~\cite{translate_DBLP:journals/corr/abs-2108-13897}. We tested using the Portuguese version of such datasets as this is the language spoken by the majority of our users. Unless otherwise noted, the experiments on MS MARCO and Robust04 uses 1000 documents from the candidate retrieval stage.

The results are shown in Table~\ref{tab:all_effectiveness_results}. Both models used in NeuralSearchX outperform BM25 by a large margin in multiple metrics and datasets. The mT5-3B reranker shows the best results for all metrics and datasets except for the MARCO dataset due to the fewer candidate documents provided to the model in that experiment. Note that while the mMiniLM model shows lower numbers than mT5-3B, it has much better efficiency, making it an attractive option to reduce costs.

\subsection{Legal Domain}

\begin{figure}
    \centering
    \includegraphics[width=1\columnwidth]{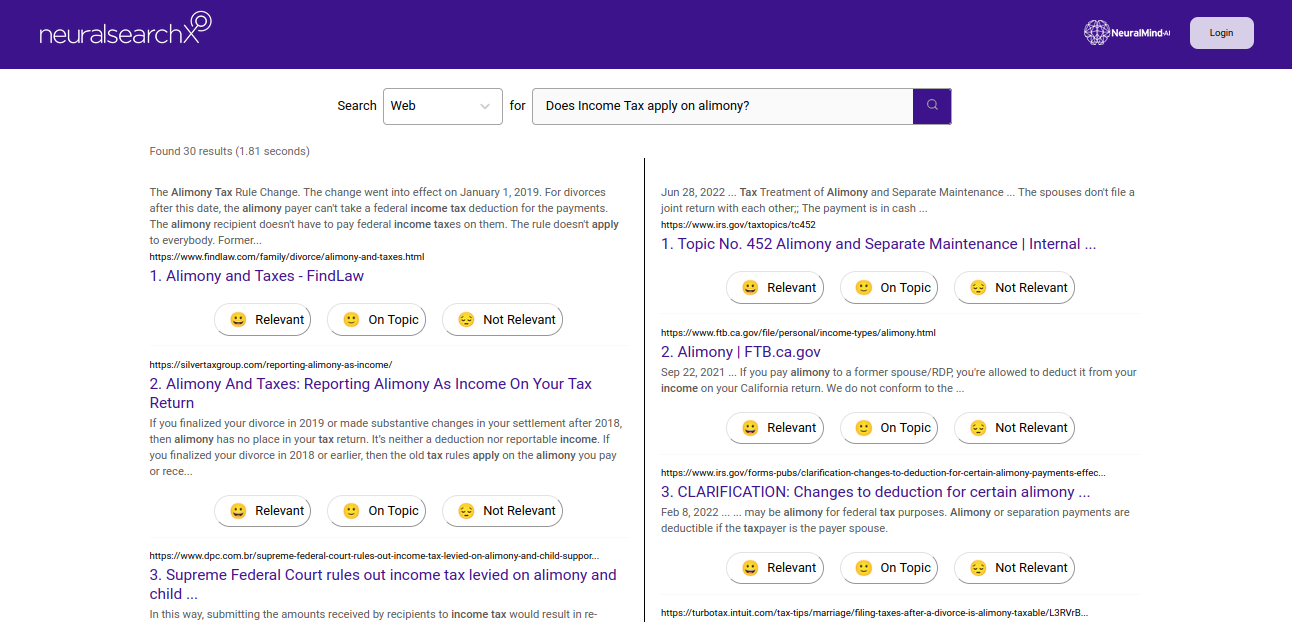}
    \caption{Annotation interface for human evaluations. Users are presented with two lists of results: one from NSX and the other from Google.}
    \label{fig:annotation_interface}

\end{figure}

To validate the quality of our results from web sources, we developed an interface in which annotators see two lists for each searched query. One of the lists displays only the Google API results in the same order that they were provided by Google, while the other list displays the results in the order given by NeuralSearchX. This list contains both Google and Bing documents. To hide the identity of the search engine, we swap the positions of the two lists with 50\% probability each time a query is issued. The top 10 results in each list are labeled by simply clicking on a button on the interface, which is shown in Figure~\ref{fig:annotation_interface}.

We then contracted attorneys to create 50 queries and annotate their results when they were issued to our system and Google Search. All queries were in Portuguese and related to Brazilian law. For each query, the attorneys labeled the first 10 documents from our search engine and the first 10 returned by Google.

Using these labels, we computed the metrics presented in column ``Legal'' of Table~\ref{tab:all_effectiveness_results}.
NeuralSearchX outperformed Google API by more than 8 nDCG@10 points.
This experiment shows that our method can surpass well-established search engines in certain domains and languages, by reranking the results obtained from third-party APIs.

\subsection{Business Domain}

Our system was selected in November 2021 to power the search engine of SEBRAE, a Brazilian entity that supports micro and small enterprises. Our system outperformed four other competitors and was chosen mainly by the quality of its results in a human evaluation.

SEBRAE provides a couple of databases for search. One of them contains long technical responses to questions from small businesses and entrepreneurs. SEBRAE receives such questions and reaches out to experts in the subject, who in turn create a technical document that answers the question in detail. Another database is from an open forum called "Sebrae Respostas" where entrepreneurs can ask anything. It works as a normal forum, but SEBRAE staff tend to answer most questions with relevant information.

We extracted the text from the provided databases to construct a common sparse index.
In addition to the databases provided, we also use Bing and Google search APIs to increase the diversity of results. We leave it to our reranker to decide which documents are shown to the end user. This system is available at \url{quest.neuralmind.ai}.

To evaluate how well our pipeline handled the documents from the business domain, we used the "technical responses" previously mentioned. Each technical response had the original question that originated the document annotated. We collected all such questions and used them as queries to evaluate the system. For each of such queries, the only document that was considered relevant was the one from which the question was extracted.

Our first result on this dataset showed a very high MRR for both NeuralSearhX and BM25.

To make the tests more discriminatory, we reduced the query set to those that performed poorly in the previous experiment by either search method. We ended up with approximately 600 test queries.

We show the results for our more discriminatory test in Table~\ref{tab:all_effectiveness_results}, column ``Business + MRR@10''.

Moreover, we performed a manual graded relevance annotation, i.e., the documents are marked by humans as non-relevant, on topic or relevant, on a curated set of 25 queries provided by SEBRAE. The annotation was performed on the top 3 results from the Google API, BM25 and NeuralSearchX. The results are shown in Table~\ref{tab:all_effectiveness_results}, column ``Business + nDCG@3''. NeuralSearchX clearly outperforms both BM25 and the Google API by a large margin. 

\section{Conclusion}

In this work we presented NeuralSearchX, a metasearch engine that takes advantage of the latest advancements in pretrained deep learning models as well as low-reliability infrastructure to make the system affordable. We demonstrated its efficiency through numerous experiments, even when using billion-parameter machine learning models. In particular, we demonstrated
competitive QPS with MS Biomedical Search while having a much lower overall cost. Finally, we showed the effectiveness of NeuralSearchX in popular public datasets and general web search, and demonstrated that it can outperform strong commercial search engines such as Google. Furthermore, we described our experience in solving and evaluating a particular use case in the business domain, showing the potential of our system for searching private collections.

\begin{acknowledgments}
This research was partially funded by grants 2020/09753-5 and 2022/01640-2 from Fundação de Amparo à Pesquisa do Estado de São Paulo (FAPESP).
We also would like to thank Microsoft Azure and Google Cloud for credits to support this work.
\end{acknowledgments}

\bibliography{main}


\end{document}